\documentclass[12pt]{article}
\usepackage{epsfig}
\usepackage{cite}
\usepackage{amsmath}
\usepackage{graphics}

\begin{document}

\setlength{\textheight}{21.5cm}
\setlength{\oddsidemargin}{0.cm}
\setlength{\evensidemargin}{0.cm}
\setlength{\topmargin}{0.cm}
\setlength{\footskip}{1cm}
\setlength{\arraycolsep}{2pt}

\renewcommand{\thefootnote}{\#\arabic{footnote}}
\setcounter{footnote}{0}

\newcommand{\gtrsim}{ \mathop{}_{\textstyle \sim}^{\textstyle >} }
\newcommand{\lesssim}{ \mathop{}_{\textstyle \sim}^{\textstyle <} }
\newcommand{\rem}[1]{{\bf #1}}
\renewcommand{\thefootnote}{\fnsymbol{footnote}}
\setcounter{footnote}{0}
\begin{titlepage}
\def\thefootnote{\fnsymbol{footnote}}

\begin{center}
\hfill arXiv:08mm.nnnn [hep-th]\\
\hfill August 2008\\
\vskip .5in
\bigskip
\bigskip

{\Large \bf Innocuous Implications of a Minimum Length in
Quantum Gravity}

\vskip .45in

{\bf Paul H. Frampton\footnote{frampton@physics.unc.edu}
}

{\it Department of Physics and Astronomy, University of North Carolina,\\
Chapel Hill, NC 27599.}

\end{center}

\vskip .4in 
\begin{abstract}
A modification to the time-energy uncertainty relation
in quantum gravity has been interpreted as increasing the duration 
of fluctuations producing virtual black holes
with masses greater than the Planck mass. I point
out that such virtual black holes have an exponential
factor arising from the action such that
their contribution to proton decay
is suppressed, rather than enhanced,
relative to Planck-mass black holes.
\end{abstract}
\end{titlepage}

\renewcommand{\thepage}{\arabic{page}}
\setcounter{page}{1}
\renewcommand{\thefootnote}{\#\arabic{footnote}}

\newpage

\vspace{3.0in}

In a remarkable paper Sakharov\cite{Sakharov} not only provided
his well-known requirements for baryogensis in the early universe
but also made the first theoretical estimate
of the proton lifetime. Ignoring all factors of order O(1),
as I shall do throughout, his formula had the form

\begin{equation}
\tau_p \sim \frac{M_{Planck}^4}{M_p^5}
\label{Sakharov}
\end{equation}
and gives a value $\tau_p \sim 10^{45}$ y. This contains
only the contribution of quantum gravity and the lifetime
in Eq. (\ref{Sakharov}) is far too long for practical measurement.
The present emprical lower bound\cite{PDG2006} on $\tau_p$ is only
$\tau_p \sim 5\times 10^{33}$y.

\bigskip

Although Sakharov did not use the language of spacetime foam,
the Euclidean path integral approach to quantum gravity provides
the means to make a similar estimate of the proton lifetime \cite{KanePerry}.  
The amplitude for producing a black hole through a fluctuation of
the space-time metric leads to a density of Planck-size black
holes of order one, in dimensionless units.  The estimate for
the proton decay rate arises from looking at virtual black
holes (hereafter VBHs) with masses in the vicinity of the Planck
mass. VBHs underly, from this viewpoint, the proton lifetime
formula found by Sakharov as in Eq.(\ref{Sakharov}).

\bigskip

To arrive at Eq.(\ref{Sakharov}) employing parallel 
methods to insights in \cite{KanePerry},
there is, in general, an exponential tunneling factor\cite{Frampton1976}. When the VBH has a 
mass close to the Planck mass, the
exponential factor is of order one, O(1), and so
does not survive in Eq.(\ref{Sakharov}).

\bigskip

For the case of VBHs which are significantly heavier than
the Planck mass, $M_{VBH} = \eta M_{Planck}$ with
$\eta \gg 1$ there was recently an interesting paper\cite{BambiFreese}
which merits further study. It arrives at
the following formula for the proton lifetime
\begin{eqnarray}
\tau_p & \sim & \frac{M_{Planck}^4}{M_p^5} \left( \frac{1}{\eta} \right)^4
\label{BambiFreese}
\end{eqnarray}
which suggests that, if $\eta$ is sufficently increased
the proton becomes disastrously unstable.

\bigskip

However, I believe upon closer scrutiny that using ideas
parallel to, although not contained in, \cite{KanePerry},
the appropriate generalization of Eq.(\ref{Sakharov}) to $\eta \gg 1$
is slightly but crucially different from Eq.(\ref{BambiFreese}) as follows
\begin{equation}
\tau_p \sim \frac{M_{Planck}^4}{M_p^5} \left( \frac{1}{\eta} \right)^4
\exp ( 4 \pi [\eta^2 - 1])
\label{FR}
\end{equation}
so there is an additional exponential factor in Eq.(\ref{FR}) for $\eta\gg 1$. 

\bigskip

The derivation of Eq.(\ref{FR}) follows from steps already implicit
in \cite{KanePerry}, especially in the Appendix thereof,
where one can find what is essentially my tunneling factor
(although it is not presented as such) in the form
\begin{equation}
\Sigma_{N=0}^{N=\infty} \exp \left[ 4 \pi N \eta^{2 + n} \right]
\label{KanePerry}
\end{equation}
where $N$ is the number of black holes and
$n$ is the number of extra dimensions.
For the present calculation, I put $N=1$ and $n=0$
in Eq.(\ref{KanePerry}). The remaining factor
in Eq.(\ref{FR}), $\exp ( 4 \pi)$, is simply to normalize 
my proton lifetime such that
there is agreement with the Sakharov formula, Eq.(\ref{Sakharov}),
in the limit $\eta \rightarrow 1$.

\bigskip

Eq.(\ref{FR}) differs from
the Sakharov formula Eq.(\ref{Sakharov}) 
for all $\eta  > 1$ by the function $f(\eta)$ where
$f(x) = \frac{1}{x^4} \exp( 4 \pi [ x^2 - 1])$.
For all $x>1$ this factor is monotonically increasing 
so that the value of the additional factor in Eq.(\ref{FR})
relative to Eq.(\ref{Sakharov}) 
for a VBH above the Planck mass, {\it i.e.}
with $\eta > 1$, leads always
to an increase in the proton lifetime,  $\tau_p$,
and there is no danger\footnote{With regard to the archived v3
of \cite{BambiFreese} this point still holds.}
of a conflict with the experimental lower limit.

\bigskip
\bigskip
\bigskip
\bigskip

\vspace{3.0in}

\begin{center}

\section*{Acknowledgement}

\end{center}

I acknowledge useful discussions with Ryan Rohm. This work was supported in part 
by the U.S. Department of Energy under Grant
No. DG-FG02-06ER41418.

\end{document}